\documentclass[cits]{PoS}
\usepackage{graphicx}
\usepackage{amsmath}
\usepackage{dcolumn}

\title{AGN jet physics from measurements of the frequency-dependent position of the VLBI radio core }

\ShortTitle{AGN Jet Physics }

\author{\speaker{Shane P. O'Sullivan}, Denise C. Gabuzda\\%
       Department of Physics, University College Cork, Ireland\\
       E-mail: \email{shaneosullivan@physics.ucc.ie}, \email{gabuzda@physics.ucc.ie}}

\abstract{Accurate measurement of the frequency-dependent shift of the self-absorbed radio core is required for multi-frequency analysis of VLBI data since absolute positional information is lost as a result of phase self-calibration. We use the cross-correlation technique of Croke \& Gabuzda (2008) on the optically thin jet emission to align our VLBA images. Our results are consistent with those obtained from the phase-referencing method, as well as alignment by model-fitted optically thin jet components. Physical parameters of the compact jet regions, such as the magnetic field strength ($B$) and the distance of the radio core to the jet origin ($r$), can be calculated from these measurements. For the source Mrk 501, we find a magnetic field strength of $0.15\pm0.04$ G in the 8.4-GHz core at a distance of $0.8\pm0.2$ pc from the base of the jet. By extrapolating our 4.6 to 15.4 GHz results for BL Lac (2200+420), we estimate magnetic field strengths of the order of 1~G in the millimetre VLBI core. Using our core-shift measurement between 1.6 and 4.8 GHz for 1803+784, we find $B_{\rm{core}}(4.8$~GHz$)=0.11\pm0.02$ G and $r_{\rm{core}}(4.8$ GHz$)=20\pm5$ pc. The phase-referencing observations of this source at 8.4 and 43 GHz by Jim\'enez-Monferrer et al.~(2008) imply $B_{\rm{core}}(43$ GHz$)=1.0\pm0.4$~G and $r_{\rm{core}}(43$ GHz$)=2.0\pm0.9$ pc.}

\FullConference{The 9th European VLBI Network Symposium on The role of VLBI in the Golden Age for Radio Astronomy and EVN Users Meeting\\
                September 23-26, 2008\\
                Bologna, Italy}

\begin{document}

\section{Introduction}
Highly collimated jets of relativistic plasma are considered to be generated and propelled outwards from the central regions of Active Galactic Nuclei (AGN) by magnetic forces surrounding the supermassive black hole, see \cite{meierjapan} and references therein. High resolution VLBI observations can detect the synchrotron emission from these jets in the radio regime at distances of the order of $10^3-10^7$ $R_S$, where $R_S$ is the Schwarzchild radius of the black hole, corresponding to the sub-parsec to parsec scales of the jet \cite{lobanovevn}. Current theoretical models \cite{meierscience2001, vlahakiskonigl2004} invoke rotation of the magnetosphere around the compact central object which creates a stiff helical magnetic field that expels and collimates the jet flow. At a distance of several hundred gravitational radii from the central engine, the natural pinching action of the magnetic field can over-collimate the flow resulting in a ``collimation shock'' which can either partially or completely disrupt the helical field structure in the jet. Recent observational evidence suggests that this ``collimation shock'' may occur in the millimetre-wave VLBI core  \cite{marschernature2008}. A significant number of VLBI observations at cm-wavelenghts support the presence of helical magnetic fields in the parsec scale radio jet (e.g., \cite{osullivan2008} and references therein), indicating that the initially ordered magnetic field is not completely disrupted and that either remnants of the earlier field structure remain or possibly a current driven helical kink instability is generated \cite{nakamurameier2004, nakamurajapan, careyhepro}.

The radio cores of AGN are generally flat-spectrum, self-absorbed regions. Since VLBI resolution is usually not sufficient to completely resolve the true optically thick radio core, the core region contains emission from around the optical depth ($\tau$) $=1$ surface and some contribution from the optically thin inner jet. From the jet model of \cite{blandfordkonigl1979}, the frequency dependence of the self-absorbed core is described by $r\propto\nu^{-1/k_{r}}$, where $r$ is the distance from the central engine and $k_r=((3-2\alpha)m+2n-2)/(5-2\alpha)$ with $m$ and $n$ describing the power-law fall-off in the magnetic field strength and particle number density, respectively, with distance from the central engine. For equipartition between the jet particle and magnetic field energy densities, the quantity $k_r=1$, with the choice of $m=1$ and $n=2$ being reasonable (e.g., \cite{konigl1981, huttermufson1986, lobanov1998}), making $k_r$ independent of the spectral index. Therefore, in equipartition, the core-shift ($\Delta r=|r_{\nu_{1}}-r_{\nu_{2}}|$) between two frequencies ($\nu_{2}>\nu_{1}$) is directly proportional to $(\nu_{2}-\nu_{1})/(\nu_{2} \nu_{1})$.
Hence, for accurate analysis of multi-frequency VLBI maps, we must correct for the frequency-dependent position of the self-absorbed radio core. 

\section{Alignment Method \& Results}
The frequency dependent core-shift is generally obtained by aligning model-fitted optically thin jet components \cite{lobanov1998}, but this can be rather difficult for jets that lack distinct components across all frequencies. Phase-reference observations \cite{marcaideshapiro1984} can be used to obtain the absolute position of the radio core by comparing the position of the target source with respect to a reference source, which should ideally be a point source, but this is a very calibration-intensive approach.

Aligning the images via the positions of optically thin jet components can be very difficult due to the complicated total intensity structure of VLBI jets, especially at low frequencies where often there are no obvious, common optically thin features in the diffuse emission region. However, using a cross-correlation technique for aligning the optically thin emission from the jet, see \cite{crokegabuzda2008} for details, we have been able to find reliable measurements for the frequency dependent core-shift of a number of sources across a wide range of frequencies.

\subsection{Mrk 501}
We have found core-shift measurements for the well studied source Mrk 501 (1652+398) from multi-frequency VLBA observations \cite{croke2005} obtained in May 1998. For this source, we are able to compare our alignment results with those based on the alignment of individual model-fitted jet components. Using space-VLBI observations of Mrk 501 at 1.6 GHz and ground-based observations at 5 GHz in April 1998, \cite{giroletti2004} were able to align these two images by aligning the positions of the optically thin jet components. This was possible due to the similarity in resolution of the 5 and 1.6 GHz images provided by the space baselines at 1.6 GHz. Our observations of Mrk 501 at 1.6 and 5 GHz could not be aligned in this manner due to the lack of distinct jet components in the poorer resolution 1.6 GHz image. Instead, we used the cross-correlation technique to align the images and found a core-shift of $0.78\pm0.07$ mas. This value compares very well with the measured core-shift of $0.72\pm0.28$ mas from the aligned images of \cite{giroletti2004}.

\subsection{BL Lac}
BL Lac (2200+420) has a smooth, extended jet structure lacking a bright distinct jet knot, across a wide range of frequencies (e.g., \cite{osullivan2008}), with which to align the images using the model-fitted component method. However, the VLBA jet images of this source are straightforwardly aligned using the cross-correlation technique. We have analysed the core-shift using VLBA observations at 4.6, 5.1, 7.9, 8.9, 12.9 \& 15.4 GHz. The excellent frequency coverage provides a degeneracy of core-shift measurements with different frequency combinations reinforcing each other. The magnitude of the separation between the 4.6 and 15.4 GHz core is $0.44\pm0.07$ mas. The direction of the shift is $16.3\pm0.9^{\circ}$ NE, which is consistent with the inner jet direction at 43 GHz \cite{osullivan2008} of $197\pm1^{\circ}$ ($180^{\circ}+17^{\circ}$) at this epoch.

\subsection{1803+784}
Using VLBA observations of the AGN 1803+784 at 1.6 and 5 GHz, we derived a core-shift of $1.27\pm0.35$ mas using the cross-correlation technique. We are able to indirectly test the accuracy of our result from independent phase-referencing observations for this source \cite{jimenez2007}. They obtained a core-shift of $0.27\pm0.13$ mas between 8.4 and 43 GHz from their astrometric analysis for this source. Using our lower frequency measurement and assuming equipartition between the jet particle and magnetic field energy densities, we predict a shift of 0.29 mas between 8.4 and 43 GHz, consistent with the phase-referencing results. Conversely, this provides some support for the assumption that equipartition holds for the compact jet region of 1803+784 between 1.6 and 43 GHz. However, it should be noted that the value of the core-shift may change at different epochs due to nuclear flares \cite{kovalev2008}.

\begin{figure}
 \begin{minipage}[t]{7cm}
 \begin{center}
 \includegraphics[width=7.0cm,clip]{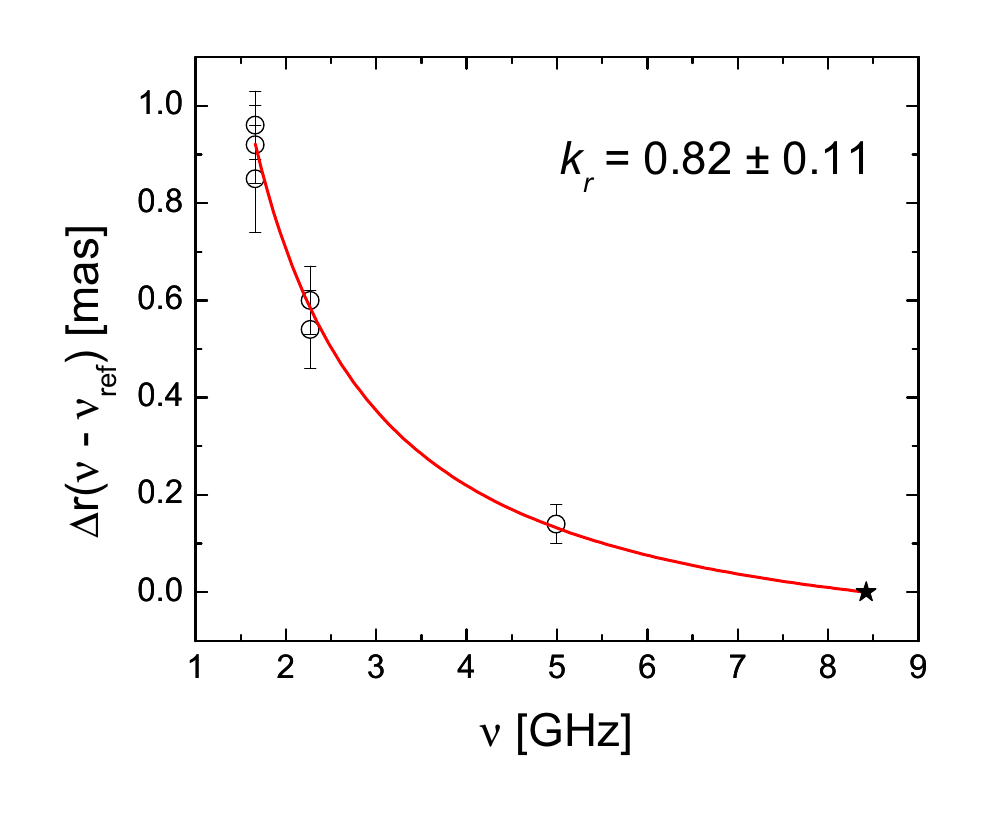}
 \caption[Short caption for figure 1]{\label{labelFig1} Plot of core-shift versus frequency for Mrk 501 using 8.4 GHz as the reference frequency. Red line: $\Delta r=A(\nu^{-1/k_r}-8.4^{-1/k_r})$ with best-fit parameters of $A=1.98\pm0.21$ and $k_r=0.82\pm0.11$}
 \end{center}
 \end{minipage}
 \hfill
 \begin{minipage}[t]{7cm}
 \begin{center}
 \includegraphics[width=7.0cm,clip]{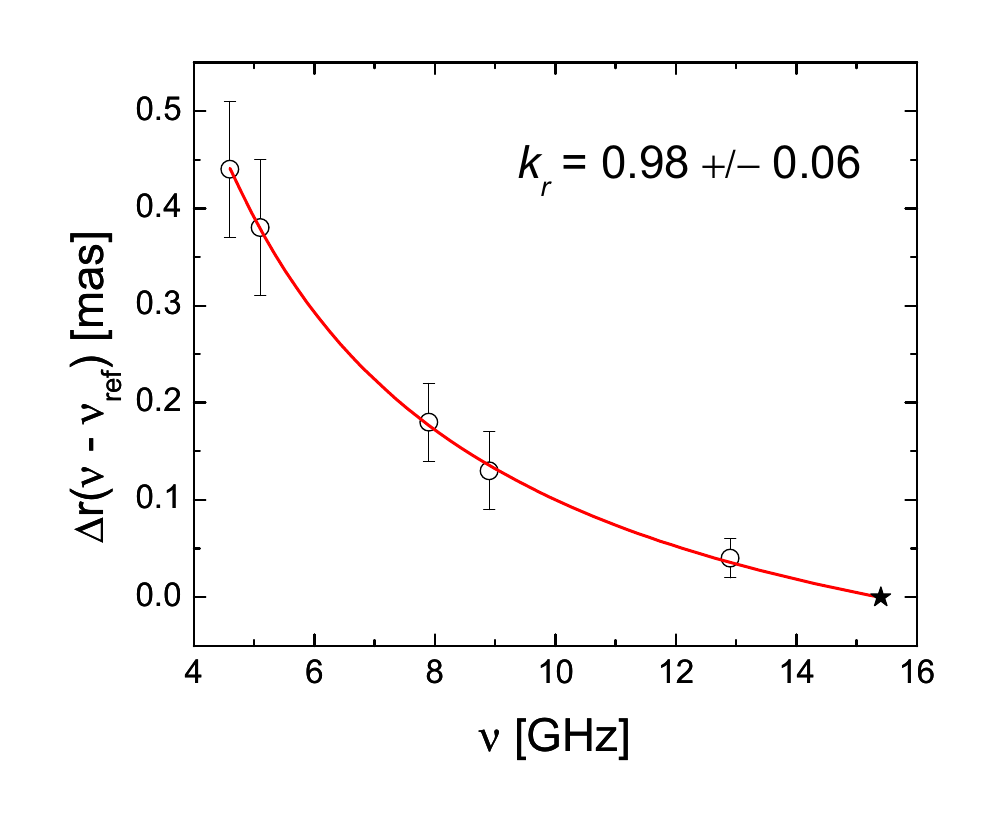}
 \caption[Short caption for figure 1]{\label{labelFig3} Plot of core-shift versus frequency for BL Lac using 15.4 GHz as the reference frequency. Red line: $\Delta r=A(\nu^{-1/k_r}-15.4^{-1/k_r})$ with best-fit parameters of $A=2.97\pm0.19$ and $k_r=0.98\pm0.06$}
 \end{center}
 \end{minipage}
 \end{figure}
 
\section{Jet Physics}
We use the equations described in \cite{lobanov1998}, from the model of \cite{konigl1981}, in order to calculate some physical parameters of the jets listed above using the core-shift measurement. Knowledge of the redshift of the source as well as estimates of the jet opening angle, viewing angle and bulk Lorentz factor are required to obtain estimates of the magnetic field strength and other quantities such as the distance of the radio core to the base of the jet.

The parameter $k_r$, described in Section 1, can be calculated from core-shift measurements at several different frequencies. Therefore, instead of assuming the equipartition value of $k_r=1$, we can measure the exact value. In the case of Mrk 501, we analysed the core-shift from observations at 1.6, 2.2, 4.8 and 8.4 GHz. Using 8.4 GHz as the reference frequency, we obtain a value of $k_r=0.82\pm0.11$ (Figure 1). This result is consistent with the synchrotron self-compton (SSC) models of \cite{katarzynski2001}, who find best fit values of $m=0.9$ and $n=1.8$ for the jet of Mrk 501, which gives a value of $k_r=0.86$ using an optically thin jet spectral index of $\alpha=-0.7$.
Using the well constrained jet parameters from \cite{giroletti2004}, we find a magnetic field strength of $0.15\pm0.04$ G in the 8.4-GHz core at a de-projected distance of $0.8\pm0.2$ pc from the jet origin. 

\begin{figure}
 \begin{minipage}[t]{7.5cm}
 \begin{center}
 \includegraphics[width=7.5cm,clip]{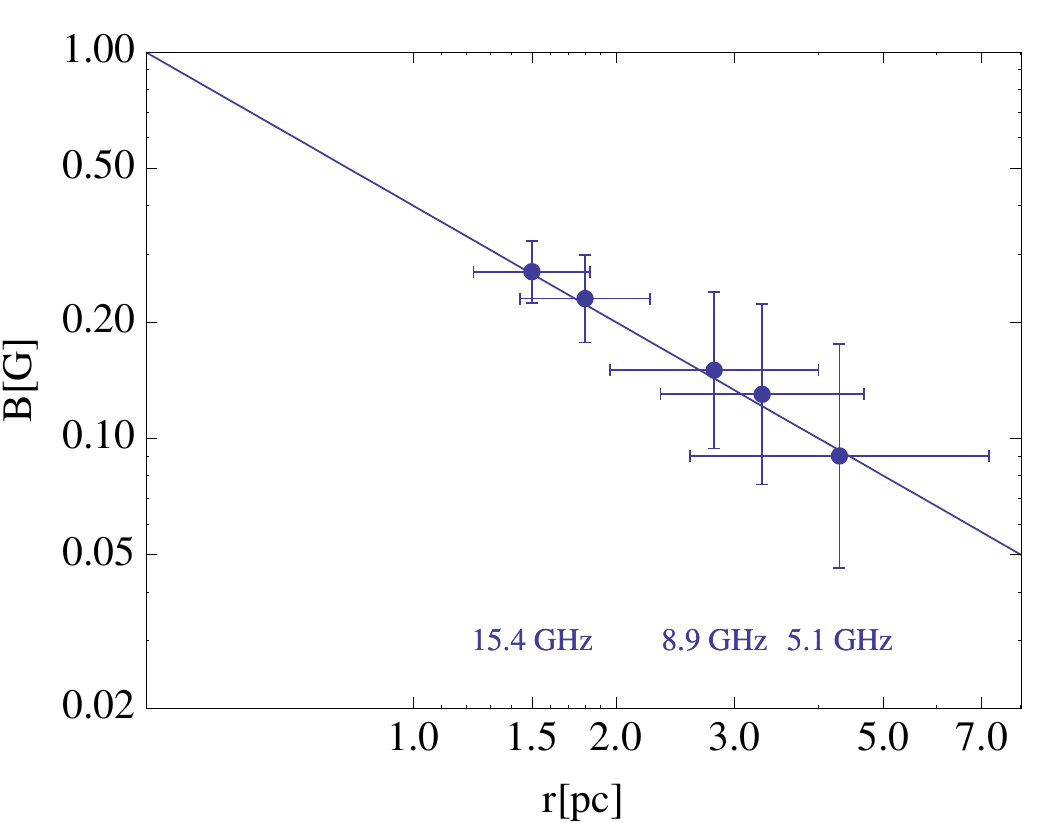}
 \caption[Short caption for figure 3]{\label{labelFig3} Plot of magnetic field strength ($B$) in units of Gauss versus distance along the jet ($r$) in units of parsecs for the core of BL Lac at 5.1, 7.9, 8.9, 12.9 and 15.4 GHz. Also plotted is the line $B=B_{1\rm{pc}}r^{-1}$ with $B_{1\rm{pc}}=0.4$ G.}
 \end{center}
 \end{minipage}
 \hfill
 \begin{minipage}[t]{7cm}
 \begin{center}
 \includegraphics[width=7.0cm,clip]{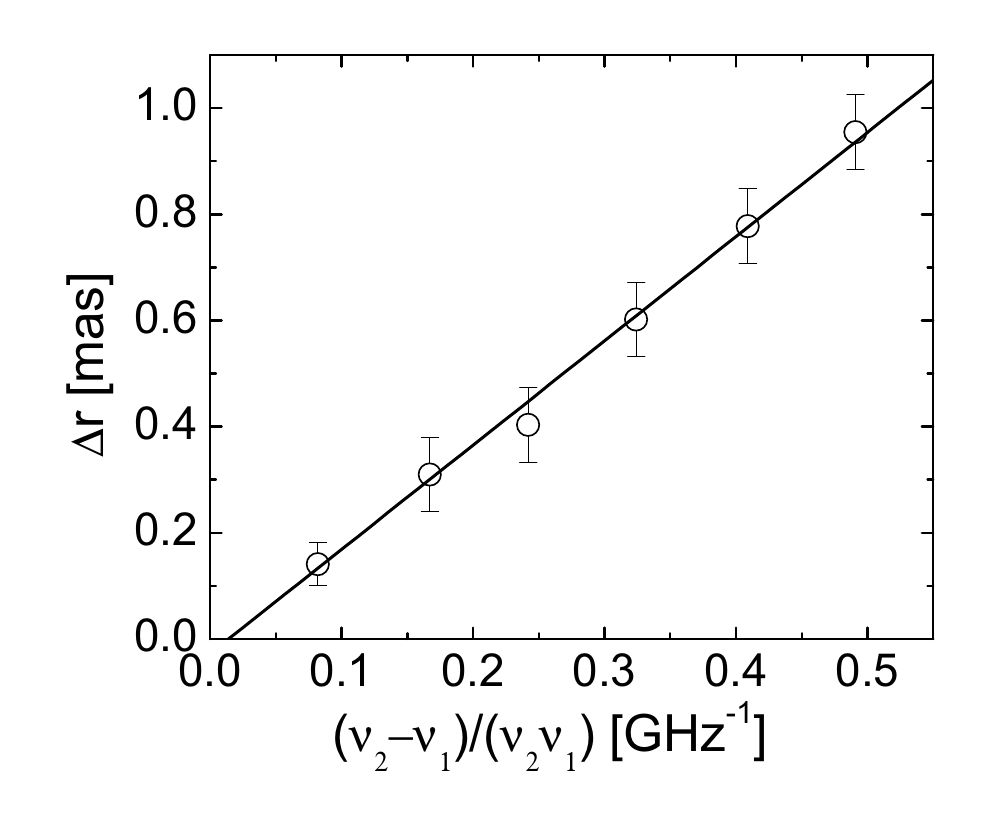}
 \caption[Short caption for figure 4]{\label{labelFig4} Straight-line fit to the core-shift data for Mrk 501 assuming equipartition. The y-intercept value is equal to $-0.03\pm0.02$ mas. A straight-line fit through the origin is expected for $k_r$ exactly equal to 1.}
 \end{center}
 \end{minipage}
 \end{figure}

We perform a similar analysis for BL Lac using observations from 4.6 to 15.4 GHz. In this case, we find a value of $k_r=0.98\pm0.06$ consistent with the equipartition value of $k_r=1$ (Figure 2). Using the jet parameters from \cite{jorstad2005}, we estimate the magnetic field strength in the VLBI core at each frequency and its distance from the base of the jet. Figure 3 shows a plot of the magnetic field strength at various distances along the jet. Extrapolating this plot, we find that the magnetic field strength approaches $\sim$1 G in the mm-core. This is consistent with results from \cite{savolainenbologna2008}, where he finds magnetic field strengths of order 1 G in the mm-core of 3C~273 by measuring the synchrotron self-absorption (SSA) turnover frequency of individual model-fitted components \cite{marscher1987}. Importantly, this method does not require the assumption of equipartition. 

Even though we have not found a value of $k_r$ for 1803+784, our results in Section 2.3 suggest that an equipartition value of $k_r=1$ is not unreasonable. Hence, we find $B_{\rm{core}}(4.8$ GHz$)=0.11\pm0.02$ G and $r_{\rm{core}}(4.8$ GHz$)=20\pm5$ pc using jet parameters from \cite{hovatta2008}. The phase-referencing measurement of \cite{jimenez2007} implies a magnetic field strength in the 43-GHz core of $1.0\pm0.4$ G at a distance of $2.0\pm0.9$ pc from the base of the jet.

One of the limitations of the core-shift method is the need to assume equipartition in order to find an estimate for the particle number density and hence, the magnetic field strength. In the case of Mrk 501, the value of $k_r$ doesn't deviate too strongly from the equipartition value (see Figure 4) but, for this source, estimating the magnetic field strength using the SSA turnover method may provide more reliable estimates. At least in the case of 2200+420, we know that the compact jet region is in the equipartition regime between 4.6 and 15.4 GHz; therefore, we have confidence that our magnetic field estimates are accurate. For 1803+784, by combining our measurements with phase-referencing observations, we have indirect evidence that the compact jet region does not deviate strongly from the equipartition regime between 1.6 and 43 GHz.

\section{Conclusions}
Using a cross-correlation technique for aligning the optically-thin jet emission, we obtain accurate measurements of the frequency-dependent shift of the self-absorbed VLBI core of three AGN jets. We have tested our method against both phase-referencing observations and alignment by individual model-fitted steep-spectrum jet components.
Using our measurements of the core-shift, we find magnetic field strengths in the compact centimetre jet regions of Mrk 501, BL Lac and 1803+784 of the order of 100's of mG; with values approaching 1 G in the mm-wave VLBI cores of BL Lac and 1803+784. We find that equipartition is a valid assumption for the compact inner jet regions of BL Lac, while we find some deviation from equipartition in the case of Mrk 501. We have also found estimates for distances of the observed VLBI core to the base of the radio jet for each source, providing the approximate location of the central supermassive black hole. 
Estimates such as these are essential in efforts to test the predictions of current theoretical models on the locations of regions of jet launching, acceleration and collimation. 

\newpage
\section{Acknowledgements}
Funding for this research was provided by the Irish Research Council for Science, Engineering and Technology. This work has also benefited from research funding from the European Community's sixth Framework Programme under RadioNet R113CT 2003 5058187.


\begin{thebibliography}{99}
  \bibitem{meierjapan} D.~L.~Meier, 2009, in Y.~Hagiwara,  E.~Fomalont,  M.~Tsuboi,  Y.~Murata, \emph{Approaching Micro-Arcsecond Resolution with VSOP-2: Astrophysics and  Technology.} ASP Conf. Ser., San Francisco, \emph{in press}
  \bibitem{lobanovevn} A.~P.~Lobanov, 2006, \emph{8th European VLBI Network Symp.}, Proc. of Science, \pos{PoS(8thEVN)003}
  \bibitem{meierscience2001} D.~L.~Meier,  S.~Koide, Y.~Uchida, 2001, \emph{Science}, {\bf 291}, 84
  \bibitem{vlahakiskonigl2004} N.~Vlahakis,  A.~K\"onigl, 2004, \emph{ApJ}, {\bf 605}, 656
  \bibitem{marschernature2008} A.~P.~Marscher, \emph{et al.,} 2008, \emph{Nature}, {\bf 452}, 966
  \bibitem{osullivan2008} S.~P.~O'Sullivan,  D.~C.~Gabuzda, 2008, \emph{MNRAS}, \emph{accepted}, {\tt arXiv:0811.4426}
  \bibitem{nakamurameier2004} M.~Nakamura,  D.~L.~Meier,  2004, \emph{ApJ}, {\bf 617}, 123
  \bibitem{nakamurajapan} M.~Nakamura,  2009, in Y.~Hagiwara,  E.~Fomalont,  M.~Tsuboi,  Y.~Murata,  eds, \emph{Approaching Micro-Arcsecond Resolution with VSOP-2: Astrophysics and Technology.} ASP Conf. Ser., San Francisco, \emph{in press}
  \bibitem{careyhepro} C.~Carey, C.~Sovinec, S.~Heinz,  2008, in \emph{High Energy Phenomena in Relativistic Outflows. International Journal of Modern Physics D}, Vol. 17, p. 1707
  \bibitem{blandfordkonigl1979} R.~D.~Blandford, A.~K\"onigl,  1979, \emph{ApJ}, {\bf 232}, 34
  \bibitem{konigl1981} A.~K\"onigl,  1981, \emph{ApJ}, {\bf 243}, 700
  \bibitem{huttermufson1986} D.~J.~Hutter, S.~L.~Mufson,  1986, \emph{ApJ}, {\bf 301}, 50
  \bibitem{lobanov1998} A.~P.~Lobanov,  1998, \emph{A\&A}, {\bf 330}, 79
  \bibitem{marcaideshapiro1984} J.~M.~Marcaide, I.~I.~Shapiro,  1984, \emph{ApJ}, {\bf 276}, 56
  \bibitem{crokegabuzda2008} S.~Croke,  D.~C.~Gabuzda,  2008, \emph{MNRAS}, {\bf 386}, 619
  \bibitem{croke2005} S.~Croke, P.~Charlot, D.~C.~Gabuzda, H.~Sol, 2005, in J.~Romney and M.~Reid, eds, \emph{Future Directions in High Resolution Astronomy: The 10th Anniversary of the VLBA}, ASP Conference Proceedings, Vol. 340, p.177
  \bibitem{giroletti2004} M.~Giroletti, \emph{et al.,}  2004, \emph{ApJ}, {\bf 600}, 127
  \bibitem{jimenez2007} S.~Jim\'enez-Monferrer,  J.~M.~Marcaide,  J.~C.~Guirado,  I.~Mart\'i-Vidal, 2008, in \emph{Modern Radio Universe: From Planets to Dark Energy.} Proc. of Science, \pos{PoS(MRU)103}
  \bibitem{kovalev2008} Y.~Y.~Kovalev, A.~P.~Lobanov, A.~B.~Pushkarev, J.~A.~Zensus, \emph{A\&A}, {\bf 483}, 759
  \bibitem{katarzynski2001} K.~Katarzy\'nski, H.~Sol,  A.~Kus,  2001, \emph{A\&A}, {\bf 367}, 809
  \bibitem{jorstad2005} S.~G.~Jorstad, \emph{et al.,}  2005, \emph{AJ}, {\bf 130}, 1418
  \bibitem{savolainenbologna2008} T.~Savolainen,  2008, \emph{these proceedings,} \pos{PoS(9thEVN)009}
  \bibitem{marscher1987} A.~P.~Marscher,  1987, in J.~A.~Zensus,  T.~J.~Pearson,  eds, \emph{Superluminal Radio Sources.} Cambridge University Press, p.~280 
  \bibitem{hovatta2008} T.~Hovatta, E.~Valtaoja, M.~Tornikoski, A.~L\"ahteenm\"aki, 2008, \emph{A\&A}, \emph{accepted}, {\tt arXiv:0811.4278}
  

\end{thebibliography}


\end{document}